\documentclass{article}

\usepackage[nonatbib, final]{nips_2017_2}

\usepackage[utf8]{inputenc} 
\usepackage[T1]{fontenc}    
\usepackage{hyperref}       
\usepackage{url}            
\usepackage{amsmath}
\usepackage{booktabs}       
\usepackage{amsfonts}       
\usepackage{nicefrac}       
\usepackage{microtype}      
\usepackage[pdftex]{graphicx}

\title{From Multimodal to Unimodal Webpages for Developing Countries}

\author{
  Vidyapu~Sandeep \qquad\qquad\qquad V~Vijaya~Saradhi \qquad\qquad\qquad Samit~Bhattacharya\\
  \texttt{s.vidyapu@iitg.ac.in} \qquad\qquad \texttt{saradhi@iitg.ac.in} \qquad\qquad \texttt{samit@iitg.ac.in}\\
   Department of Computer Science and Engineering\\
   Indian Institute of Technology Guwahati\\
   Assam, India 781039 \\
}

\begin{document}

\maketitle

\begin{abstract}
The multimodal web elements such as text and images are associated with inherent memory costs to store and transfer over the Internet. With the limited network connectivity in developing countries, webpage rendering gets delayed in the presence of high-memory demanding elements such as images (relative to text). To overcome this limitation, we propose a Canonical Correlation Analysis (CCA) based computational approach to replace high-cost modality with an \textit{equivalent} low-cost modality. Our model learns a common subspace for low-cost and high-cost modalities that maximizes the correlation between their visual features. The obtained common subspace is used for determining the low-cost (text) element of a given high-cost (image) element for the replacement. We analyze the cost-saving performance of the proposed approach through an eye-tracking experiment conducted on real-world webpages. Our approach reduces the memory-cost by at least 83.35\% by replacing images with text.
\end{abstract}

\section{Introduction}
World Wide Web (WWW) serves the basic information needs of people across the globe. Information access is provided through the Internet while the presentation is through multimodal elements text, image, audio and video as rendered on web interfaces. These modalities though help to present the information in multiple ways, they differ in the inherent costs of storage and transmission over the Internet. Thus, the information rendering on the webpages gets affected by the Internet bandwidth available to the users. That is, for a given bandwidth, webpages with low-cost elements (text) gets rendered quicker than the equivalent webpage with the high-cost elements (images). Even though several optimization techniques are proposed to reduce the storage and transmission costs of the  modalities, such as image compression methods, the inherent costs cannot be reduced beyond a threshold. 

The Internet bandwidth is limited in the developing and least developed countries (LDCs) when compared with the developed countries. As per ICT Facts and Figures~\cite{sanou2017ict}, in the year 2016, the international Internet bandwidth per user is 6 kilobits per second (kbps) for LDCs and 53 kbps for developing countries while it is 140 kbps for the developed countries. It means, for example, to visualize a complete 7.5kB (or 60 kb) image on the webpage, users from LDCs have to wait for 10 seconds. Such high latency causes user dissatisfaction besides affecting their psychological and cognitive processing. Thus, apart form other solutions that address limited bandwidth problem, web designers can choose to render the content with low-cost modalities to serve the information needs of the progressive 41.3\% of people from developing countries (Internet users)~\cite{sanou2017ict}. However, the unimodal conversion of the multimodal webpages, through replacement of high-cost modalities with the low-cost modalities, should achieve equivalence from two representational aspects of the modalities: 1. \textit{semantic features}--- to represent the semantic information associated with the elements, 2. \textit{visual features}--- to represent how users visually perceive the elements presented on the interface. Towards this, the semantic cross-modal analyses (given a text fetch the semantically equivalent image and vice-versa)~\cite{wang2016comprehensive} helps to obtain the equivalent semantic low-cost modality of a given modality. However, such equivalent modality cannot be used for direct replacement as user attention needs to be preserved even after the replacement. The preservation is desired for both the user and the website owner as web elements are designed so as to draw user's desired attention.

We propose to replace high-cost modality (image) with the low-cost modality (text) by obtaining the equivalent low-cost elements based on visual attention, analogous to semantics based cross-modal fetching in~\cite{rasiwasia2010new}. For this, we pair the text and image elements based on their associated visual attention. Then, a common subspace is learned (called \textit{correlated subspace}) that maximizes the correlation between paired elements using canonical correlation analysis (CCA)~\cite{hotelling1936relations}. The common subspace thus achieved is used for procuring the equivalent low-cost element for the replacement. In the following section, we describe the pairing of text and image elements based on their associated attention.

\section{Cross-Modal Pairing based on Visual Attention}
Eye-tracking is used to measure the visual attention directly~\cite{zaphiris2007human}. The \textit{fixations}--- gazing towards a location for certain threshold of time; indicates the attention. The elements that are fixated are called attended elements or fixated elements. The fixations are ordered to give \textit{fixation indices} (FIs) to the fixated elements. That is, the first fixated element obtains FI of one, second fixated element obtains FI of two and so on. Thus, FIs indicate the attention drawing ability of web elements in the order of their associated indices. So, we propose to pair attended cross-modal elements based on their FIs as
\begin{align}~\label{eq:clustMedia}
D = \Big\{\big(T_{i}, I_{i}\big) \quad | \quad \mathcal{FI}^{T_{i}}_{w} = \mathcal{FI}^{I_{i}}_{w} \quad \forall w \in W\Big\}
\end{align}
where, $(T_{i}, I_{i})$ is the pair of text and image elements whose FIs respectively $\mathcal{FI}^{T_{i}}$ and $\mathcal{FI}^{I_{i}}$ on the webpage $w$ are matched. $W$ is the set of webpages considered for analysis and $D$ is the dataset of paired text and images.

\section{CCA based Computational Approach}
\begin{figure}[h]
  \centering
   \includegraphics[height=0.3\linewidth]{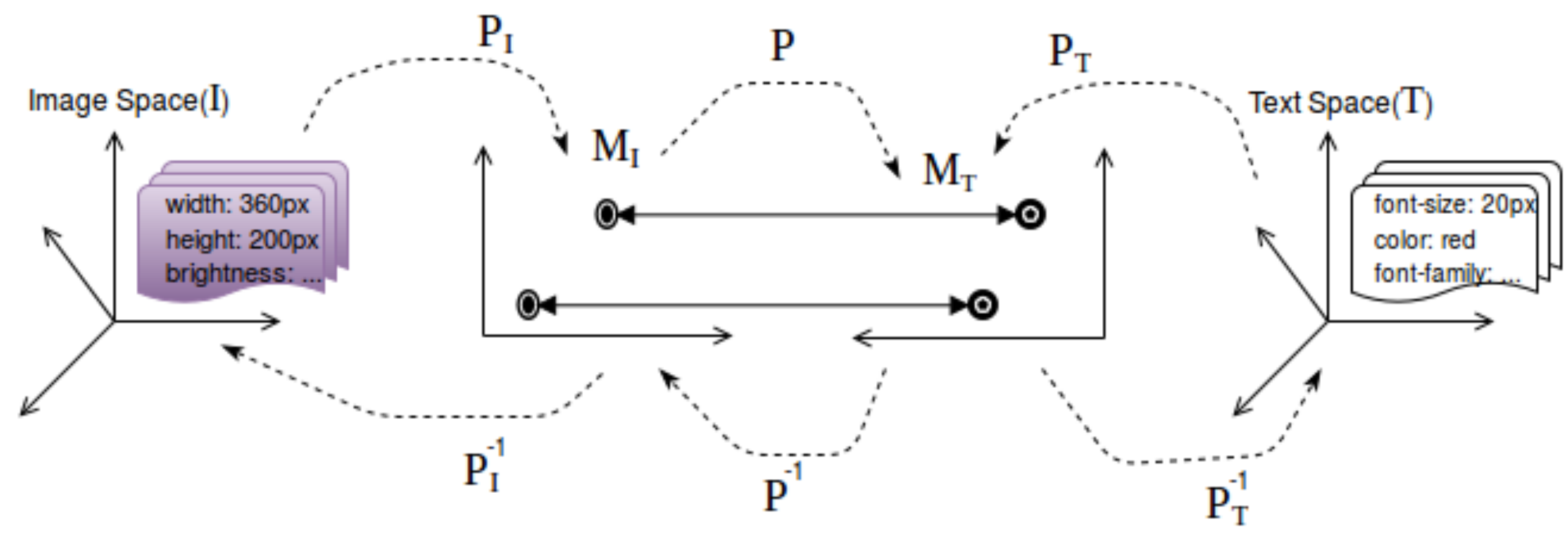}
  \caption{Illustrative diagram of obtaining one modality (say `T') element from the other (say `I') using the projection matrices ($\text{P}_\text{T}$, $\text{P}$ and $\text{P}_\text{I}$) and maximally correlated subspaces (`$\text{M}_\text{T}$' and `$\text{M}_\text{I}$'). Feature spaces `T' and `I' are constructed from visual features of text and images respectively.}~\label{fig:CCA}
\end{figure}
Each paired text element $T_i$ is considered as a data point in text space `T' and each paired image element $I_i$ is a data point in the image space `I'. As there is no natural correspondence between `T' and `I', data points are projected onto the subspaces that have correspondence between them. Subspaces are obtained by projecting data along directions that maximizes correlation between paired points. This is the classical formulation of canonical correlation analysis (CCA).  Using CCA~\cite{hardoon2004canonical, hotelling1936relations}, the projection matrices $\text{P}_\text{T}$ and $\text{P}_\text{I}$ are obtained to project the text and image data respectively onto the correlated subspaces `$\text{M}_\text{T}$' and `$\text{M}_\text{I}$' as shown in figure~\ref{fig:CCA}. The projection directions, that is, the columns of $\text{P}_\text{T}$ and $\text{P}_\text{I}$ are the eigenvectors of $\Sigma_{TT}^{-1/2}\Sigma_{TI}\Sigma_{II}^{-1}\Sigma_{IT}\Sigma_{TT}^{-1/2}$ and $\Sigma_{II}^{-1/2}\Sigma_{IT}\Sigma_{TT}^{-1}\Sigma_{TI}\Sigma_{II}^{-1/2}$ respectively. Here, $\Sigma_{TI}$ denotes the covariance matrix between points from `T' space and `I' space. Then, to obtain the attentionally equivalent text element of a given image query $I_q$, the nearest neighbor of $\text{P}_\text{T}^{-1} \times \text{P} \times \text{P}_\text{I} \times I_q$ (that is $I_q$ projected into `T' through `$\text{M}_\text{I}$' and `$\text{M}_\text{T}$') is obtained from the text space `T'. The equivalent text element thus obtained is used for replacing the corresponding image element on the webpage. To analyze the performance of this approach, we conducted an eye-tracking experiment as described.

\section{Eye-tracking Experiment}
\paragraph{Stimuli and Experimental Setup:} We selected 22 real-world webpages ($|W| = 22$) belonging to sports, news, Wikipedia, e-commerce and education for the stimuli. This selection helps to overcome the influence of webpage template bias on the user attention. Each webpage is displayed in full-screen mode and the screenshot is captured on a 22" display monitor with $1680 \times 1050$ resolution. 
A slideshow presentation was prepared using these webpages with a blank slide inserted after every webpage to reset the attention of the user. The display duration of each webpage was chosen five seconds as we are interested in visual features than the high-level semantic features. The slides were presented in the \textit{Counterbalanced} mode to avoid the ordering effects of webpages on participants' attention. The slideshow was presented on a 22" desktop monitor with a screen resolution of $1680 \times 1050$ using Tobii Pro software. Tobii X2-60 Eye-tracker {\url{https://www.tobiipro.com/product-listing/tobii-pro-x2-60/}} was used for capturing the participants' visual attention while they gaze to the stimuli. 

\paragraph{Participants and Procedure:}
A total of 43 Indian college students (22 male and 21 female) participated in our eye-tracking experiment. Their age ranges from 21 years to 34 years with a mean of 24.3 years (SD = 3.2 years). All participants were experienced and active web users with a mean experience of 8.6 years. 
Participants took part in the experiment one at a time. For each participant eye-tracker was calibrated using the typical 9-point method to sync participant's eye-movements with the tracker. After successful calibration, prepared automated slideshow was presented which the participant has freely-viewed. During the presentation, eye-tracker captured the gazing behavior of the participant.

From the captured gaze-data, fixations with a gaze duration of at least 100 milliseconds were extracted. The fixation locations are used to traverse the document object model (DOM)~\url{https://www.w3.org/TR/WD-DOM/introduction.html} of the respective webpage to identify the fixated element. The fixated elements' attributes are extracted to prepare the visual features for analysis. 

\section{Visual Features}~\label{subsec:visual}
We considered the visual features of the text and images that are shown to influence the attention of the users. As the position of element influences the attention, the positional distances of the element from four edges of the screen are considered as visual features besides its area.
\paragraph{Text:} The cascading style sheets (CSS)~\url{https://www.w3.org/Style/CSS/} attributes that influence the visual appearance of the text elements on the webpage are extracted. From these attributes visual features are extracted such that (1) each color attribute contributed four visual features corresponding to red, green, blue color components along with the opacity of the color and (2) each space-related attribute (margin and padding) contributed four visual features corresponding to the space on top, left, right and bottom sides of the text element. In total 70 text visual features are considered. 
\paragraph{Image:} Wide range of applications were proposed based on Itti's model~\cite{itti1998model} for attention on images. Thus, we considered the color and intensity related visual features for the analysis. The color histograms were computed for each of red, green and blue color components. Each color component contributed eight visual features corresponding to the uniform distribution of color component values that range from 0 to 255. Similarly, as some applications used gray-scale values to account for the intensity of the images, we also computed the eight uniformly spaced histograms for the gray color. Other image specific visual features include brightness, luminance, hue, saturation, value (HSV), mean and spread (or variance) of red, green and blue color components. We also computed the contrasting features of the images (subtracting image features from the respective features computed for the webpage as images) to account for their saliency on the webpage. In total 91 image visual features are considered for the analysis.

The extracted visual features are paired as per equation~\ref{eq:clustMedia}. This resulted in $|D| = 14330$ paired data points. Then, the CCA based computational approach is applied on these paired data points to learn a common subspace. Given an image, \textit{attentionally} equivalent text is determined using the common subspace that is used for the replacement on the corresponding webpage. The cost-saving analysis of the replacement is described in the following section.

\section{Result and Analysis}
Our approach achieved a very high correlation of 0.9948 between the visual features of text and images in the common subspace with 28 dimensions. The high-correlation demonstrates the usefulness of common subspace in replacing images with text on webpages whose cost-saving analysis is presented in rest of the section.

There were 139 images on the selected stimuli of 22 webpages with an average of 6.32 images per webpage. The total cost (in terms of memory storage) of these images is 8573.05 kB with an average image cost of 61.68 kB. Thus the images induce an average cost of 389.8 kB per webpage. When the whole screen (resolution: $1680 \times 1050$) is filled with medium-sized text (font-size: 16 px), it can accommodate $(1680 \times 1050) / (16 \times 16) \approx 6890$ characters. As web-browsers use up to 4 Bytes per character in Unicode representation, the text cost is 27560 Bytes (or 26.91 kB). Including the additional cost of 6890 Bytes to format the text on webpage, the total text cost is 33.64 kB. With these costs involved, we present the cost-saving and thereby webpage rendering time-saving for the developing countries (with 53 kbps~\cite{sanou2017ict}) as follows.

(1) \textbf{Minimum cost-saving} occurs when a single image that occupies whole screen is replaced with equivalent text element. The percentage of cost saved is $(61.68 - 33.64)\times 100/33.64 = 83.35\%$. (2) \textbf{Maximum cost-saving} occurs when the screen-full of images are replaced with text occupying the screen. Then the percentage of cost-saved is  $(389.8 - 33.64)\times 100/33.64 = 1058.74\%$. (3) \textbf{Achieved cost-saving:} 
We used fixation indices (FIs) of the query image element and the obtained text element to analyze the replacement performance as they are indicative of human attention. As FIs ranged from 1 to 23 with the latter FIs being sparse, we considered the FIs up to the median of FIs for the analysis. The replacement performance (that is, micro-F1 score) is 0.52. Thus, the average cost saved is $(389.8 \times \mathbf{0.52} - 33.64)\times 100/33.64 =  502.54\%$. In this way, the webpage that takes 30.6 seconds (for $389.8 \times 0.52 = 202.696$ kB) to render the information, now takes 5.07 seconds (for 33.64 kB) to render the equivalent low-cost information.
In summary, our approach helps in designing attention-preserving low-cost modality webpages that serve the 41.3\% of people from developing countries.

\small
\bibliographystyle{unsrt}
\bibliography{nips1_bibliography}

\end{document}